\documentclass[conference]{IEEEtran}
\usepackage{cite}

\ifCLASSINFOpdf
\usepackage[pdftex]{graphicx}
\else
\fi
\usepackage{amsmath}
\usepackage{amsfonts} 
\usepackage{amsthm}
\usepackage{amssymb}
\usepackage[boxruled]{algorithm2e}
\usepackage{amsthm} 
\usepackage{mathtools} 
\usepackage{textcomp} 

\usepackage{BOONDOX-ds}

\usepackage{bbm}

\newcommand{\pfun}{\mathop{\hbox{$\to$\kern-7pt\raise.9pt\hbox{\scalebox{1}[.55]{$|$}}\kern4pt} }}

\usepackage{array}

\begin{document}

\title{AI Data Wrangling with Associative Arrays}

\author{\IEEEauthorblockN{Jeremy Kepner$^{1,2,3}$, Vijay Gadepally$^{1,2}$, Hayden Jananthan$^{1,4}$, Lauren Milechin$^5$, Siddharth Samsi$^1$
\\
\IEEEauthorblockA{$^1$MIT Lincoln Laboratory Supercomputing Center, $^2$MIT Computer Science \& AI Laboratory, \\ $^3$MIT Mathematics Department, $^4$Vanderbilt University, $^5$MIT Department of Earth, Atmospheric and Planetary Sciences}}}
\maketitle

\begin{abstract}
The AI revolution is data driven.  AI ``data wrangling'' is the process by which unusable data is transformed to support AI algorithm development (training) and deployment (inference).  Significant time is devoted to translating diverse data representations supporting the many query and analysis steps found in an AI pipeline.  Rigorous mathematical representations of these data enables data translation and analysis optimization within and across steps.  Associative array algebra provides a mathematical foundation that naturally describes the tabular structures and set mathematics that are the basis of databases.  Likewise, the matrix operations and corresponding inference/training calculations used by neural networks are also well described by associative arrays.  More surprisingly, a general denormalized form of hierarchical formats, such as XML and JSON, can be readily constructed.  Finally, pivot tables, which are among the most widely used data analysis tools, naturally emerge from associative array constructors.  A common foundation in associative arrays provides interoperability guarantees, proving that their operations are linear systems with rigorous mathematical properties, such as, associativity, commutativity, and distributivity that are critical to reordering optimizations.
\end{abstract}

%
\IEEEpeerreviewmaketitle

\section{Introduction}
\let\thefootnote\relax\footnotetext{This material is based upon work supported by the Assistant Secretary of Defense for Research and Engineering under Air Force Contract No. FA8702-15-D-0001 and National Science Foundation CCF-1533644. Any opinions, findings, conclusions or recommendations expressed in this material are those of the author(s) and do not necessarily reflect the views of the Assistant Secretary of Defense for Research and Engineering or the National Science Foundation.}

AI models require data for their construction and training \cite{lecun2015deep}.  The broad applicability of AI to diverse domains involves integration and conditioning of diverse data.  AI ``data wrangling'' is the process by which unusable data is transformed to support AI algorithm development (training) and deployment (inference) \cite{mckinney2012python}.  Significant time is devoted to translating diverse data representations supporting the many query and analysis steps found in an AI pipeline \cite{AIdata2018}.  Typical steps include parsing data from a raw hierarchical format to approximately tabular spreadsheet files, ingesting into databases, querying for specific AI analysis, construction of  vectors for training models, and storing matrix representations of neural networks.  Each of these steps can be represented using a plethora of data structures and formats and a major component of AI data wrangling is harmonizing these representations to align with the different operations of an AI pipeline.

Theory can play a significant role in harmonizing data representations.  Set theory forms the basis of the mathematical duality enabling the declarative SQL user interface to be implemented with any procedural language in a relational database \cite{codd1970relational, maier1983theory, codd1990relational}.  Graph theory and linear algebra are the basis of the matrix representation of AI neural networks \cite{kepner2011graph, 7761646, 8091098, 8392693}.  Associative array algebra generalizes set theory and matrices to further unify SQL, NoSQL, and NewSQL databases \cite{gadepally2015graphulo, kepner2016associative, kepnerjananthan}.  Polystore databases leverage associative array mathematics to simplify the interchange of data among diverse databases \cite{7761636, mattson2017demonstrating, jananthan2017polystore}.

Hierarchical data structures, such as JavaScript Object Notation (JSON) \cite{crockford2006application} and Extensible Markup Language (XML) \cite{bray2000extensible}, are important to AI pipelines.  Spreadsheets, and their corresponding pivot tables, are available to over 100 million users and are an important tool for presenting AI results.   This work presents rigorous mathematical representations of these data as associative arrays, enabling data translation and analysis optimization within and across AI pipeline steps. 

\section{Associative Array Mathematics}
\begin{figure*}[]
\centering
\includegraphics[width=7in]{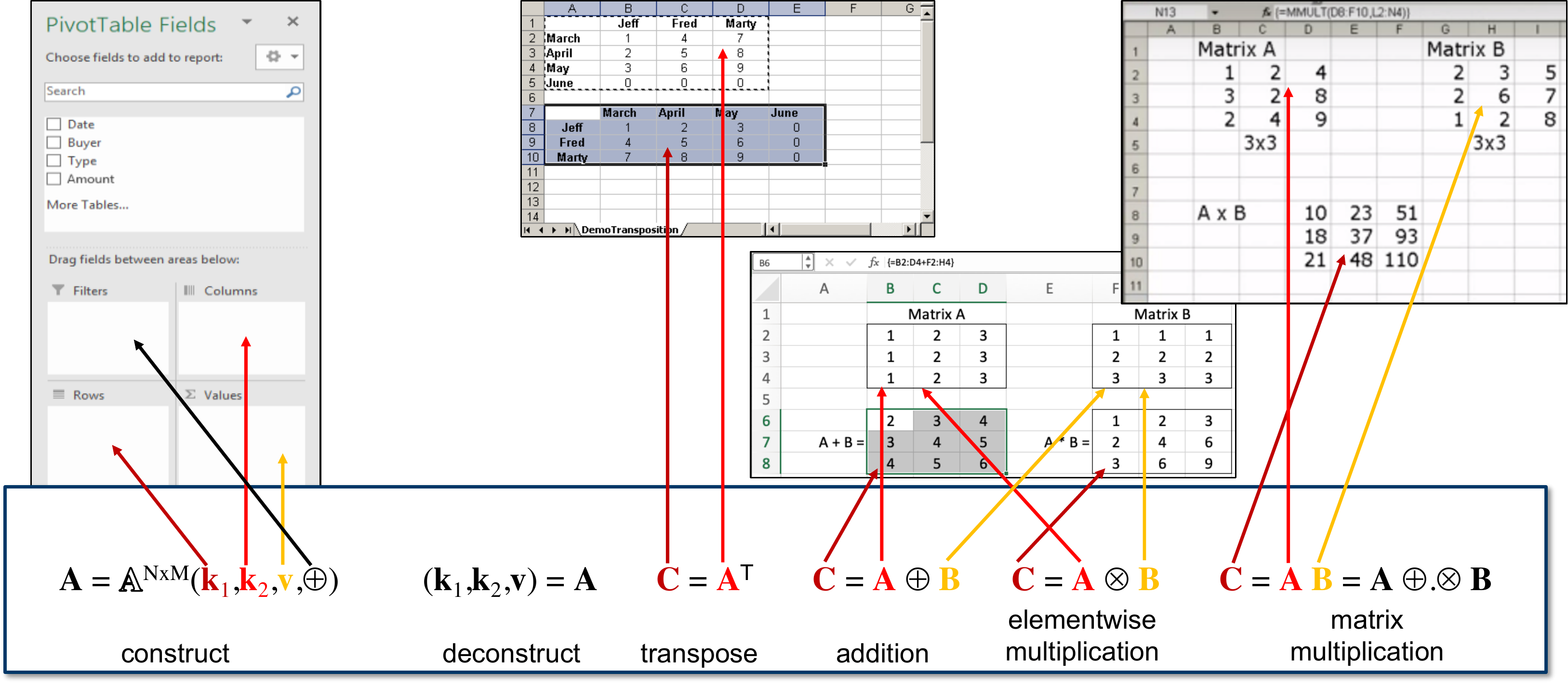}
\caption{Core associative array operations and their corresponding spreadsheet operations depicted in Microsoft Excel \cite{ExcelPivotTable, ExcelTranspose, ExcelAddition, ExcelMatrixMultiplication}.}
\label{fig:PivotTable}
\end{figure*}

\begin{figure*}[]
\centering
\includegraphics[width=7in]{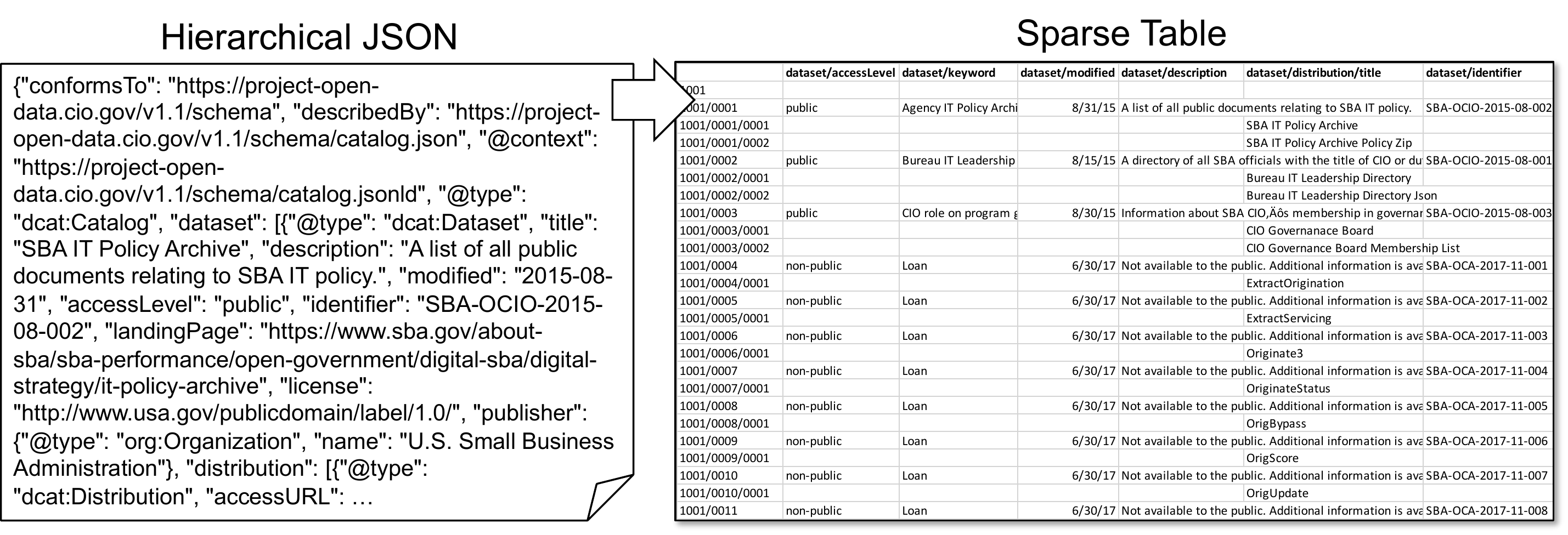}
\caption{Hierarchical JSON metadata records from Data.Gov Small Business Administration  \cite{DataGovSBA} and their corresponding denormalized sparse representation.}
\label{fig:HierarchicalSparse}
\end{figure*}

  The full mathematics of associative arrays and the ways they encompass matrix mathematics and relational algebra are described in the aforementioned references \cite{jananthan2017polystore,kepner2016associative,kepnerjananthan}. The essence of associative array algebra is three operations: element-wise addition (database table union), element-wise multiplication (database table intersection), and array multiplication (database table transformation).  These operations are illustrated as spreadsheet operations in Figure~\ref{fig:PivotTable}. In brief, an associative array $\mathbf{A}$ is defined as a mapping from sets of keys to values
$$
  \mathbf{A}: K_1 \times K_2 \to \mathbb{V}
$$
where $K_1$ are the row keys and $K_2$ are the column keys and can be any sortable set, such as sets of integers, real numbers, and strings. The row keys are equivalent to the sequence ID in a relational database table.  The column keys are equivalent to the column names in a database table.  $\mathbb{V}$ is a set of values that forms a semiring $(\mathbb{V},\oplus,\otimes,0,1)$ with addition operation $\oplus$, multiplication operation $\otimes$, additive identity/multiplicative annihilator 0, and multiplicative identity 1. The values can take on many forms, such as numbers, strings, and sets. One of the most powerful features of associative arrays is that addition and multiplication can be a wide variety of operations.  Some of the common combinations of addition and multiplication operations that have proven valuable are standard arithmetic addition and multiplication ${+}.{\times}$, union and intersection ${\cup}.{\cap}$ used in database operations, and various tropical algebras that are important in finance and neural networks: ${\max}.{+}$, ${\min}.{+}$, ${\max}.{\times}$, ${\min}.{\times}$, ${\max}.{\min}$, and ${\min}.{\max}$.  Because associative arrays are typically constructed as semirings their operations are linear systems with rigorous mathematical properties, such as, associativity, commutativity, and distributivity that are critical to reordering optimizations.

\section{AI Pipelines}

Hierarchical data structures, such as JSON and XML, are playing an increasingly important role in AI pipelines because of their ability to capture diverse data of the type that lends itself to AI applications.  JSON and XML also align well with object oriented programming as arrays of objects built from other arrays objects can be written in a human readable format.  However, most data analysis environments and AI pipelines depend upon data being in an approximately tabular format.  Fortunately, hierarchical data can also be represented as sparse associative arrays by traversing the hierarchy and incrementing/appending the row counters and column keys to emit row/column/value triples directly passable to an associative array constructor (see Figure~\ref{fig:HierarchicalSparse}).  As sparse associative arrays hierarchical data can be readily ingested into AI pipelines.

  Spreadsheets are used by over 100 million people each day.  The results of AI pipelines are often presented in spreadsheet form.  Figure~\ref{fig:PivotTable} illustrates the spreadsheet equivalents of the core associative arrays operations.  Interestingly, pivot tables, which are among the most widely used spreadsheet analysis tools, naturally emerge from associative array constructors.

 Associative arrays span the data and operations found in most AI pipelines, providing a powerful tool for harmonizing and optimizing AI pipeline data flow.

\section*{Acknowledgement}

The authors wish to acknowledge the following individuals for their contributions and support: Bob Bond, Alan Edelman, Laz Gordon, Charles Leiserson, Dave Martinez, Mimi McClure, Michael Wright, William Arcand, David Bestor, William Bergeron, Chansup Byun, Matthew Hubbell, Michael Houle, Michael Jones, Anna Klein, Peter Michaleas, Julie Mullen, Andrew Prout, Antonio Rosa, Charles Yee, and Albert Reuther.


%
%



\bibliographystyle{ieeetr}

\bibliography{AIdataWrangling}
%

\end{document}